\documentclass[12pt]{article}
\usepackage{algpseudocode}
\usepackage{amsmath, amssymb, amsthm}
\usepackage{float, fullpage, graphicx, multirow,parskip, subcaption, setspace}
\usepackage{comment}
\usepackage{url}
\usepackage{enumitem}
\usepackage{tikz}
\usepackage{bm, bbm}
\usetikzlibrary{shapes.geometric, positioning}
\usetikzlibrary{quotes, angles}
\usepackage{rotating}
\usepackage{hyperref}
\usepackage{natbib}
\usepackage{placeins}
\usepackage{algorithm2e}
\usepackage{geometry}
\RestyleAlgo{ruled}
\usepackage{lineno}
\usepackage{rotating}
\usepackage{booktabs} 
\usepackage{longtable}
\usepackage{tabularx,colortbl}

\definecolor{SkyBlue}{RGB}{14, 118, 188}
\definecolor{BrightRed}{RGB}{223,82, 78}

\hypersetup{pdfborder = {0 0 0.5 [3 3]}, colorlinks = true, linkcolor = BrightRed, citecolor = SkyBlue}

\def\keywordname{{\bfseries \emph Keywords}}%
\def\keywords#1{\par\addvspace\medskipamount{\rightskip=0pt plus1cm
\def\and{\ifhmode\unskip\nobreak\fi\ $\cdot$
}\noindent\keywordname\enspace\ignorespaces#1\par}}

\title{Identify local limiting factors of species distribution using min-linear logistic regression}
\author{Hongliang Bu\thanks{School of Life Sciences, Peking University}  
\and Yunyi Shen\thanks{Dept. of Electrical Engineering \& Computer Science, Massachusetts Institute of Technology, Correspondence to: yshen99@mit.edu}}

\begin{document}
\begin{spacing}{1.9}

\def\bY{\bm{Y}}
\def\by{\bm{y}} 

\def\bz{\bm{z}}
\def\bX{\bm{X}}
\def\bx{\bm{x}} 

\def\R{\mathbb{R}}
\def\N{\mathcal{N}}
\def\P{\mathbb{P}}
\def\E{\mathbb{E}}

\def\Xcal{\mathcal{X}}

\maketitle

\begin{abstract}
Logistic regression is a commonly used building block in ecological modeling, but its additive structure among environmental predictors often assumes compensatory relationships between predictors, which can lead to problematic results. In reality, the distribution of species is often determined by the least-favored factor, according to von Liebig's Law of the Minimum, which is not addressed in modeling. To address this issue, we introduced the min-linear logistic regression model, which has a built-in minimum structure of competing factors. In our empirical analysis of the distribution of Asiatic black bears (\textit{Ursus thibetanus}), we found that the min-linear model performs well compared to other methods and has several advantages. By using the model, we were able to identify ecologically meaningful limiting factors on bear distribution across the survey area. The model's inherent simplicity and interpretability make it a promising tool for extending into other widely used ecological models.

\end{abstract}
\keywords{von Liebig's law \and Species distribution modeling \and Limiting factor \and Conservation planning}

\section{Introduction}
\label{sec:introduction}

Logistic regression is one of the the working horses in ecological modeling for binary outcomes, such as presence-absence or detection-nondetection, and is frequently employed to assess species distribution. Besides directly using it as a species distribution model \citep[SDM]{pearce2000evaluation,stockwell2002effects}, it is also used in models like occupancy models \citep{MacKenzie2002, Mackenzie2006}, capture-recapture models \citep{royle2013spatial}, and various zero-inflated models \citep{agarwal2002zero,zuur2016beginner}. Logistic regression is commonly grounded in niche theory, which posits that the model can identify the boundary of a species' realized niche within a high-dimensional niche space. However, the additive structure of logistic regression implies certain implicit constraints on the geometry of the realized niche when incorporating multiple environmental predictors. A critical implicit assumption of logistic regression is the compensatory relationship among environmental predictors. When using an additive structure in the model, such as between food abundance and human population density as a proxy for disturbance, the model assumes that low human population density can compensate for low food abundance. However, a species' distribution is typically governed by the least favorable environmental condition, even if other factors are highly favorable for the species. Such observation is known as von Liebig's Law of the Minimum in population biology \citep{odum1971fundamentals,danger2008does}, which is not fully addressed though while  it is in particular needed in both theoretical and applied fields. 

Although quantifying local limiting factors in logistic regression can be challenging, several approaches have been employed to assess the impact of such factors on a species' response. Either based on expert knowledge of species environmental limits, or defining limiting factor through a roundabout way by measuring similarity between sites in estimation and training sites \citep{li2008}, such methods fail to estimate limiting factors directly but expediently. Regression quantiles model the upper bounds of species-environment relationships and tend to describe potential rather than actual patterns of species distribution. Model selection of regression quantiles is based on a range of quantiles (1-99th) that can be estimated, and some objective criteria have to be proposed to choose a single quantile after selection for the model's application \citep{Vaz2008}.
Our aim is to address the challenge of identifying limiting factors and overcoming the implicit compensatory assumption inherent in logistic regression, all while preserving the model's interpretability. To achieve this, we utilize a recently developed extension of logistic regression-based models featuring a minimal linear structure that explicitly models the limiting factor \citep{Xu2019}.  
In recent statistical research, \citet{Xu2019} introduced the concept of max-linear logistic regression as a generalization of the traditional logistic regression model. Unlike logistic regression, which models the log odds of distribution as a linear combination of predictors, max-linear logistic regression models log odds as the maximum of several linear combinations of predictors known as factors. As a result, this model has a dominating factor that is explicitly built into its formulation. While the likelihood function for such models is not smooth due to the maximum structure, it has been shown that maximum likelihood estimation is consistent and produces good results \citep{Xu2019}.

Through a comparison of the max-linear regression model with various classification methods, including Random Forest and Support Vector Machine, \citet{Xu2019} demonstrates that the max-linear regression model performs as well as, or even better than, these other methods. In the ecology context, we propose using the min-linear logistic regression model (hereafter referred to as min-linear LR), which is mathematically equivalent to the max-linear LR but has greater ecological appeal. We use this model to identify the limiting factor for species' response and to understand the significance of environmental predictors for guiding conservation practices. The min-linear LR has an intrinsic ``limiting factor'' built into the model, making it easy to identify the environmental predictor that limits the species at a certain site simply by identifying which factor reaches the minimum.

In this paper, we introduced the min-linear LR model which explicitly added a minimum structure accounting for limiting fator theory. Then we illustrated the use of the model with empirical distribution data of Asiatic black bear (\textit{Ursus thibetanus}) collected by \citet{liu2009spatial} in Sichuan Province, China. Finally, we discussed the potential of incorporating this modification in modelling procedure into the framework of widely used hierarchical models, such as occupancy models.


\section{Methods}
\label{sec:proposed_method}
\subsection{The min-linear LR}

Here we consider situations in which surveys are performed at \textit{N} discrete and independent sites on species distribution status. Let $Y_i$ denote the binary observations of presence/absence at site $i=1,2,...,N$, where possible values are $Y_i=0$ ("absence") and $Y_i=1$ ("presence") and $Y_i$ is a Bernoulli process with success probability $p_i$. Multiple environmental factors of concern $\mathbf{X_i}$ are recorded at each site. This is a typical data structure for logistic regression model which is defined as:

\begin{equation}
\label{eqn:logistic}
    \begin{aligned}
    P(Y_i=1)&=p_i=\frac{\exp\{\mathbf{X}_i\beta\}}{1+\exp\{\mathbf{X}_i\beta\}}\\
    P(Y_i=0)&=1-P(Y_i=1)
    \end{aligned}
\end{equation}

Instead of the additive structure in above equation, the min-linear LR model exploits the structure of competing factors and is defined as:

\begin{equation}
\label{eqn:minlogistic}
    \begin{aligned}
    P(Y_i=1)&=\frac{\exp(\min\{\alpha_1+\mathbf{X}_i^{(1)}\beta_1,\dots,\alpha_L+\mathbf{X}_i^{(L)}\beta_L\})}{1+\exp(\min\{\alpha_1+\mathbf{X}_i^{(1)}\beta_1,\dots,\alpha_L+\mathbf{X}_i^{(L)}\beta_L\})}\\
    P(Y_i=0)&=1-P(Y_i=1)
    \end{aligned}
\end{equation}

In the above equation, $\mathbf X_i^{(\ell)}\in \mathbb{R}^{n\times p_\ell}$ is the design matrix corresponding to the $\ell$-th factor ($1\le \ell\le L$) which is a group of independent variables connected in linear relationship, $\beta_\ell\in \mathbb{R}^{p_\ell}$ is the vector of coefficient for the $\ell$-th factor, and $\alpha_\ell$ is the intercept for the $l$-th factor, controlling importance level of the $l$-th factor. The factor matrices are allowed to have overlapping columns, but cannot be completely identical to avoid unidentifiability of parameters \citep{Xu2019}.

One challenge for fitting this model is that the likelihood function is not smooth because of the minimum structure. We could use the so-called gumbel max trick with a tuning parameter $a$ \citep{huijben2021}, i.e. 

\begin{equation}
    \sum_i\frac{\exp(-ax_i)}{\sum_i \exp(-ax_i)}x_i \to \min \{x_i\} \text{ as } a\to\infty
\end{equation}

The model then can be fitted via the method of maximum likelihood, and model selection is easily accomplished based on AIC or BIC framework.  For a set of simulations see \citet{Xu2019}. The non-smooth likelihood function in our model also precludes estimation of Wald confidence interval; instead, we can apply parametric bootstrap to quantify the uncertainty.

Based on the min-linear LR, we can make predictions of both distribution probability $p_i$ and the limiting environmental factor, i.e. the factor which makes the minimum at a site. Note that the parameters in the model can be explained as some different kinds of importance, for instance $\alpha_\ell$, if it is large compared with other $\alpha$'s and all environmental variables standardized before mode fitting, then the factor is less likely to be the minimum at a given site; while if $\beta$'s are large, small change of a certain predictor could bring about disproportional large change of the log-odds given the factor is the limiting factor (i.e. the species is sensitive to the environment when it is limiting). Finally, logistic regression is algorithmically identified as a special case of min-linear LR model if we set all but one of the intercepts $\alpha_\ell$'s to be $\infty$.  


\subsection{Distribution data of Asiatic black bear}

The presence-absence data were obtained from \citet{liu2009spatial}, which aimed to estimate distribution of Asiatic black bear in Sichuan Province, southwestern China. The province covers a larger area and assorted topography and habitats. In the eastern region are the Sichuan Basin and surrounding low mountains, and the western region is the Tibetan Plateau, between which are the Hengduan Mountains consisting of a group of north-south mountain ranges. The survey divided the province into 15 km $\times$ 15 km cells and sampled 21\% (494) of the total. Occupancy status of bears in selected cells was investigated by systematically designed survey protocol which started from interview of local villagers and are followed by confirmation through activity evidence such as claw marks, feeding platforms, etc., and searching for bear signs along transects if negative report was obtained from interview. 

\subsection{Model fitting and comparison with other models}

We analyzed limiting factors for Asiatic black bear distribution with our min-linear LR model. We considered five environmental predictors of four groups, namely conservation status, which was the percentage of a grid inside natural reserve(s), two topography measures, i.e. elevation and terrain ruggedness, human population density and forest cover percentage \citep{SHEN2021e01831} (Details see Table.\ref{tab:predictor}). All the covariates were standardized to z-score before model fitting.

We also fitted Random Forest and logistic regression models with the same dataset mentioned above, generated Asiatic black bear distribution maps, and compared the results with min-linear LR based on ROC-AUC in a 5-fold cross-validation. For the Random Forest model, which is a tree model and does not have an explicit parametric form, we calculated the Gini index to measure the importance of each variable \citep{Breiman2001}. We conducted all analysis in the program R \citep{rprogram}.

\section{Results}
\label{sec:results}

The models were fitted successfully. The parameter estimates of $\beta$s for min-linear LR and logistic regression analysis are rather different (Table \ref{tab:my_label}). The most impressive difference is that the factor human density has significantly negative impact on bear distribution in the former model, but no significant impact in the later one. Random Forest model, as a tree model, does not have an explicit parametric form, thus estimation errors for $\beta$s are not available. However, in no matter which model, conservation status is either an insignificant factor or has minimal importance (Table \ref{tab:my_label}).

\begin{table}[htb]
    \centering
    \caption{Parameter estimates (standard deviation) of min-linear logistic regression and general logistic regression models, and importance of variables measured by Gini index for Random Forest model. Intercept estimates for different feature groups were listed separately at the lower half.}
    \begin{tabular}{lccc}
    \hline
    Environmental predictors & min-linear LR & logistic regression & Gini index \\
    \hline
    \textit{Conservation} &&& \\
    \quad Conservation status & 13.97 (7.77) & 0.31 (0.68) & 6.67 \\
    \textit{Topology} &&& \\
    \quad Elevation & 0.40 (0.26) & 1.43 (0.48) & 30.78 \\
    \quad Ruggedness & 2.50 (0.47) & 1.69 (0.29) & 60.14\\
    \textit{Human} &&& \\
    \quad Human density & -8.42 (3.97) & 0.063 (0.45) & 25.47\\
    \textit{Forest} &&& \\
    \quad Forest cover & 11.37 (3.67) & 1.72 (0.35) & 37.89\\
    \hline
    Group intercepts & & &\\
    Overall & N/A & 1.01 (0.22) & N/A\\
    Conservation & 7.58 (7.14) &N/A &N/A\\
    Topology & 1.59 (0.30) &N/A & N/A\\
    Human & 13.78 (6.35) &N/A &N/A \\
    Forest & 13.76 (5.92) &N/A &N/A\\
    \hline
    \end{tabular}
    \label{tab:my_label}
\end{table}

\begin{figure}[htb]
    \centering
    \includegraphics[width = \linewidth]{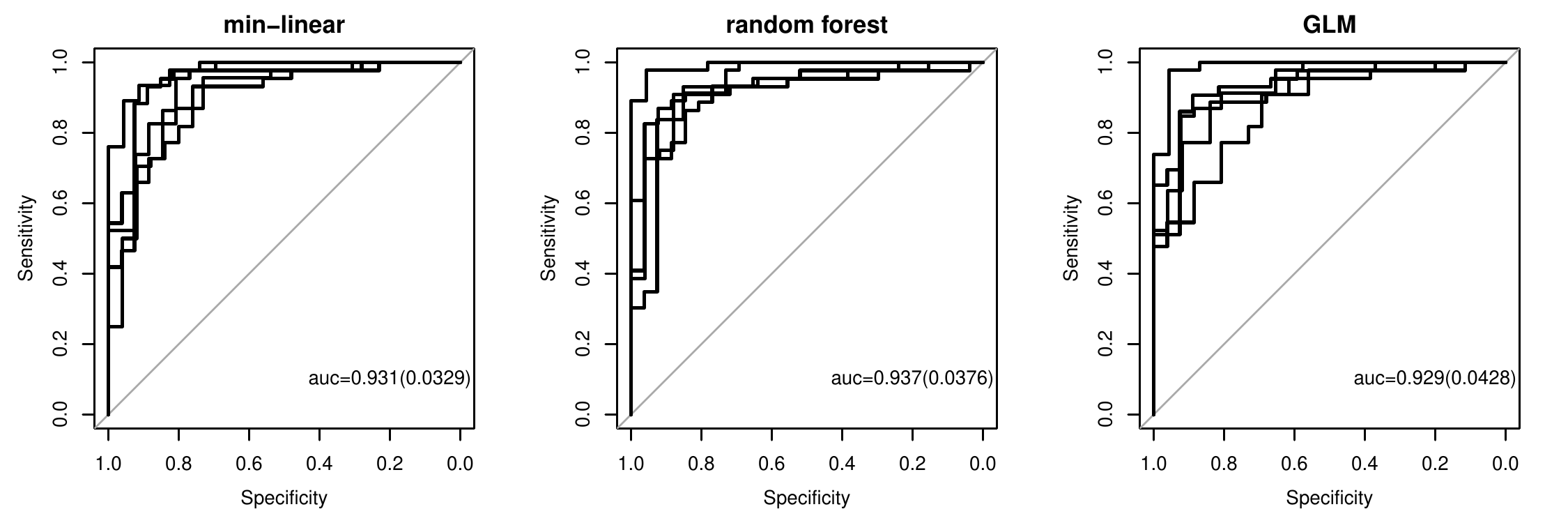}
    \caption{ROC-AUC results in the 5-fold cross validation of the min-linear logistic regression, Random Forest, and logistic regression. there is no significant difference in AUC measures among the three methods}
    \label{fig:auc}
\end{figure}

\begin{figure}[H]
    \centering
    \includegraphics[width = \linewidth]{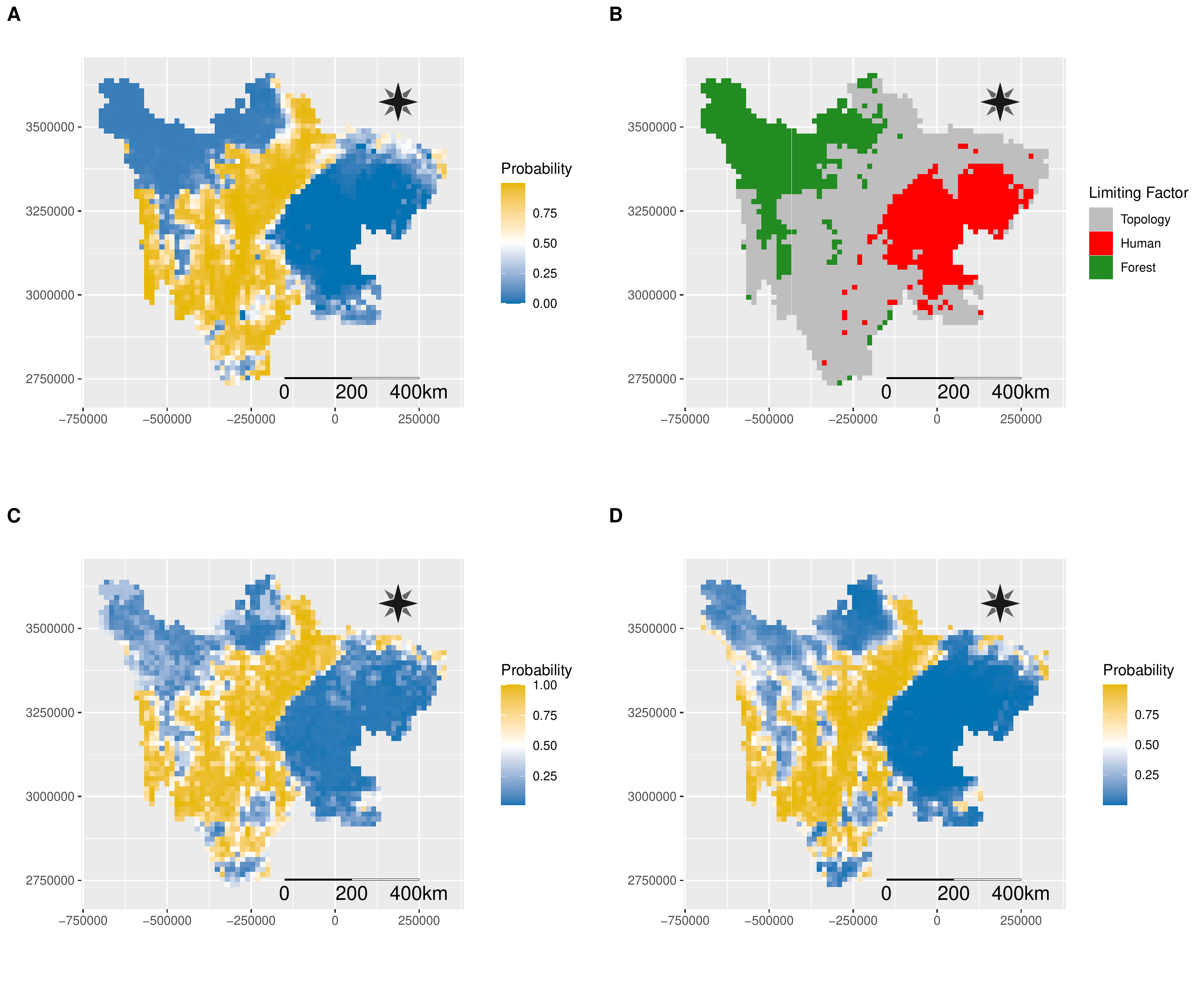}
    \caption{Distribution maps predicted by min-linear LR (A), Random Forest (C) and logistic regression (D), and limiting factors identified from min-linear LR (B).}
    \label{fig:my_label}
\end{figure}

Based on AUC values (Figure. \ref{fig:auc}), min-linear LR performed as good as Random Forest and logistic regression models. According to the model result of min-linear LR, we predicted limiting factors in all cells across the study area. We identified three main regions that had different limiting factors, namely in most of Sichuan Basin the black bears were limited by human populations, in Hengduan Mts., they were mostly limited by topology and secondly by forest cover in some cells, while in most Qinghai-Tibet Plateau it was forest cover that played the limiting role (Figure. \ref{fig:my_label}). Corresponding to modeling results, conservation status is not limiting factor in any region. Distribution probability maps predicted with all models are roughly similar, except Northwest Sichuan region, where bears are predicted absent from min-linear LR but present with low probabilities from the other two models.

\section{Discussion}
\label{sec:discussion}

By incorporating von Liebig's Law of the Minimum through a built-in minimum structure, we have extended the widely used general logistic regression model into the min-linear logistic regression model. This model performs well despite its non-smooth likelihood function. In fact, \citet{Xu2019} suggested that the model is especially capable of capturing non-linear relationships, which cannot even be handled by logistic regression. As a result, the min-linear LR could in some cases outperform some black-box machine learning models, such as Random Forest and support vector machine, while maintaining simplicity and interpretability by preserving linear relationships within feature groups.

In addition to providing a robust method for modeling species' binary responses like distribution status, the min-linear LR model has the added benefit of identifying important groups of features that are meaningful for both ecology and conservation. While other classification methods such as logistic regression, Random Forest, and support vector machine (SVM) offer measures of ``variable importance,'' these measures are typically based on information that should be interpreted as ``importance in predicting the distribution'' on a landscape scale rather than a local or site scale. In contrast, the important features identified in the min-linear LR model represent the limiting environmental factors that have local effects on a species' distribution, corresponding to local adaptation of a population and providing valuable information for conservation management. For instance, it is possible that a species is sensitive to humans (i.e. a small change in human density can incite a large amount of change in the distribution probability), but at a certain site human density is very low, and other environments, say forest cover determines the distribution. In this case, human should not be understood as an "important" factor and management aiming to lower human disturbance is obviously misleading, even though it is on the landscape level. The min-linear LR model can distinguish between these by having a large $\beta$ value for humans but in a particular site, the human factor is not necessarily attending the minimum. By comparing the predicted distribution maps of Asiatic black bears in our study with those in \citet{liu2009spatial}, we were able to demonstrate the practical advantages of the min-linear LR model. While the logistic regression model used in \citet{liu2009spatial} overestimated bear distribution in northwestern Sichuan, our min-linear LR model successfully identified forest cover as the limiting factor and provided a more accurate prediction of bear distribution.

The clear probabilistic formulation of min-linear LR made it easy to extend to several other types of models with good reason. Occupancy models are widely used to model wildlife-environment relationships while accounting for imperfect observations \citep{MacKenzie2002, Mackenzie2006} in studies like camera-trap and other repeated surveys. It is logistic regression based, and it could be an improvement for the occupancy model when the min-linear structure is incorporated, as both occupancy probability and detection probability are usually limited by specific environmental features. From here, our model can extend in various ways to be multivariate (e.g. multispecies, spatial, or temporal explicit). The min-linear structure can also be incorporated into Poisson regression models, which implies its application in Poisson regression-based models, such as N-mixture model \citep{royle2004}. Another possible application is in the Cox proportional-hazards model which is used in wildlife survival analysis such as nesting success \citep{leighton2011,nur2004}. The predictors are assumed to act additively on survival, but it is more reasonable to allow predictors to compete with each other. Although testing and use of these extensions were necessary and were out of the scope of this paper, we believe the model could make genuine contributions to the ecological study and conservation practice. 

\section*{Acknowledgement}
This research has received no external funding. The authors used ChatGPT to improve English writing. The author would like to thank Jiangyue Wang from PKU for her comments on an early draft of this manuscript. 

\bibliography{references.bib}

\section*{Appendix}
\label{sec:appendix}
\appendix
\renewcommand{\theequation}{A\arabic{equation}}
\renewcommand{\thesection}{A\arabic{section}}  
\renewcommand{\thefigure}{A\arabic{figure}}  
\renewcommand{\thetable}{A\arabic{table}}  
\setcounter{equation}{0}
\setcounter{table}{0}
\setcounter{section}{0}
\setcounter{subsection}{0}
\setcounter{subsubsection}{0}
\section{Environmental predictors}
\label{app:env}
\begin{table}[htbp]
	\caption{Sources and abbreviations of environmental predictors used to predict the potential habitat of Asiatic black bears at the coarse and fine resolutions, respectively.}
	\begin{center}
		\begin{tabular}{llll}
			\toprule
			Name (units)   & Source & Data type  \\
			
			\midrule
			Elevation (m)  & NASA SRTM\textsuperscript{a} & continuous \\
			Roughness (m)  & from Elevation& continuous \\
			Forest cover (0-1 unitless)  & Global Forest Watch\textsuperscript{b} & continuous \\
			Human density (/km$^{2}$)  & Harvard IQSS\textsuperscript{c} & continuous \\
			Conservation status (percentage protected)  & \citet{SHEN2021e01831} & continuous \\
			\bottomrule		
		\end{tabular}
	\end{center}
	Notes: \textsuperscript{a}NASA Shuttle Radar Topography Mission, \textsuperscript{b}https://www.globalforestwatch.org/, \textsuperscript{c}Harvard Institute for Quantitative Social Science
	\label{tab:predictor}
\end{table}

\end{spacing}

\end{document}